%% file: kklcc.tex
\documentclass[aps,preprint,
tightenlines 
]{revtex4}
\usepackage{pictex,graphicx,color,rotating}     
\input jek                

\long\def\beginnote#1\endnote{\begingroup\ttfamily\small#1\endgroup}
\long\def\beginhidden#1\endhidden{}

\long\def\beginnote#1\endnote{\begingroup\ttfamily\small#1\endgroup}

\begin{document}

\preprint{SNUTP 01-001,hep-th/0101027}
\title{Self-tuning solution of the cosmological constant 
problem with antisymmetric tensor field}
\author{Jihn E. Kim, Bumseok Kyae and Hyun Min Lee}
\affiliation{
Department of Physics and Center for Theoretical Physics,\\
Seoul National University, Seoul 151-747, South Korea
}
\date{\today}
\pacs{PACS: 11.25.Mj, 12.10.Dm, 98.80.Cq}

\begin{abstract}
We present a self-tuning solution of the cosmological constant problem
with one extra dimension which is curved with a warp factor.
To separate out the extra dimension and to have a self-tuning
solution, a three index antisymmetric tensor field is 
introduced with the $1/H^2$ term in the Lagrangian.
The standard model fields are located at the $y=0$ brane. 
The existence~\cite{kklcc} of the self-tuning solution 
(which results without any fine tuning among parameters in the
Lagrangian) is crucial to obtain a vanishing cosmological constant 
in a 4D effective theory. The de Sitter and anti de Sitter space
solutions are possible. The de Sitter space solutions have horizons.
Restricting to the spaces which contain the $y=0$ brane, the
vanishing cosmological constant is chosen in the most probable 
universe. For this interpretaion to be valid, the existence 
of the self-tuning solution is crucial in view of the phase
transitions. In this paper, we show explicitly a solution
in case the brane tension shifts from one to another value.
We also discuss the case with the $H^2$ term which leads to 
one-fine-tuning solutions at most.  
\end{abstract}
\maketitle

\def\p{\partial}
\def\L{\Lambda}

\section{Introduction}

The cosmological constant problem~\cite{veltman} is probably 
the most important clue to the physics at the Planck scale. 
Most attempts toward solutions of the cosmological constant
problem introduce additional 
ingredients~\cite{witten,baum,hawking,coleman,others,duff},
except the wormhole and anthropic solutions
of the problem~\cite{coleman,anthropic}.

The wormhole solution 
is based on the probabilistic interpretation that the
probability to have a universe with a vanishing 
cosmological constant is the largest~\cite{coleman}. But 
in the evolving universe where the true vacuum is chosen 
as the universe cools down, the probabilistic interpretation 
in the early universe is in question since at a later
epoch additional constant may be generated by spontaneous
symmetry breaking. For this interpretation to make sense,
there must exists a {\it self-tuning solution}. [The {\it self-tuning} 
solution is defined as the flat space solution without any
fine-tuning of parameters in the action.]
The anthropic interpretation~\cite{anthropic} 
for the small cosmological constant is a working proposal, 
but does not answer the fundamental question, $\lq\lq$Can we 
explain the vanishingly small constant from the fundamental 
parameters in the theory?" 

On the other hand, Hawking's probably vanishing cosmological
constant~\cite{hawking} 
relies on an undetermined integration constant. He showed
that the wave function for a flat universe is infinitely
large compared to non-flat universes. Thus, if there
exists an undetermined integration constant, the phase transition
will end probably to a flat universe.
However, his original proposal with a three index antisymmetric
tensor field does not introduce any dynamics in the 4D space-time
and the integration constant is just another cosmological 
constant~\cite{hawking,duff}. 
Witten~\cite{witten} also argued that the existence of 
an undetermined integration constant may be a clue to the 
understanding of the flat universe, since a many-body theory may 
provide an explanation from a one constant universe to another 
constant universe~\cite{dee}. 
Thus, in the absence of a self-tuning solution of the
cosmological constant, we regard that the 
cosmological constant problem in Hawking's scenario 
still remains as an unsolved problem.

Therefore, it is worthwhile to search for any new
solution of the cosmological constant problem. Since the
gravitational law beyond the Planck scale is not a settled
issue at present, we can look for solutions even in models
with drastically different gravity beyond the Planck scale. 
If the cosmological constant problem is solved in models
with a different gravity, then the new interaction can
be studied extensively whether it leads to another
inconsistency in theory or in phenomenology. In this spirit, 
there have been attempts to understand the cosmological 
constant in the 5 dimensional (5D) world with the fifth dimension
$y$ compactified~\cite{rs1} (RSI model) or 
uncompactified~\cite{rs2} (RSII model).
It seems that there can be a way to understand the
cosmological constant problem~\cite{shapo} in these RS type models 
since, RSI model for example, the nonvanishing brane tensions $k_1$ 
and $k_2$  and bulk cosmological constant $k$
can lead to a flat space solution if these parameters satisfy
two fine-tuning conditions $k_1=k=-k_2$. Therefore, the first
step to understand the cosmological constant is to find the
flat space solutions without any fine-tuning between parameters
in the action. 

With the fifth dimension compactified (RSI), the attempts to
solve the cosmological constant has failed so far.
The original try to find a flat space solution without any
fine-tuning~\cite{kachru} had a singularity, and taking the 
singularity into account by putting a brane there reproduced 
a fine-tuning condition~\cite{nilles} even though the two 
conditions have been reduced to just one. Therefore, the first 
step toward a solution of the cosmological constant in the RS 
type models is to have flat space solutions without any 
fine-tuning between parameters in the Lagrangian. 
Restricting just to an exponentially small cosmological
constant, it has been pointed out that it is possible
with two or more branes~\cite{hbk,tye,cline}.

Recently, the needed flat space solutions in the RSII 
type background have been found~\cite{kklcc}. With a smeared-out 
brane a similar attempt led to a self-tuning 
solution~\cite{tamvakis}.
In the RS type models, the brane(s) has a special meaning in
the sense that the matter fields can reside at the brane only.
However, the effective gravitational interaction of the matter 
fields is obtained after integrating out the fifth dimension $y$.
For this effective theory to make sense, we must require that:\\
\indent (i) the metric is well-behaved in the whole region
of the bulk, and\\
\indent (ii) the resulting 4D effective Planck mass is finite.

The condition (i) is to find a solution without a singularity
in the region defined. In this regard, if the warp factor
vanishes at say $y=y_m$, then $y_m$ becomes the horizon and the
universe connected to the matter brane is up to $y=y_m$. In the
RSI models, it is a disaster if $0<y_m<\frac{1}{2}$ is between the
two branes located at $y=0$ and $y=1/2$. This is because one 
needs both branes for the consistency in the RSI models. But in the
RSII models, if there exists $y_m$ then one can consider the
space only for $-y_m<y<y_m$ if $y_m$ is not a naked singularity.
Indeed, it can be shown that it is consistent to consider the space 
up to this point only by calculating the effective cosmological constant
by integrating $y$ to $y_m$.
The localized gravity condition (ii) restricts the solutions severely,
since the $y$ integration in some solutions would give a divergent 
quantity for $M_P$ or $M_P$ get more important 
contributions as $y\rightarrow\infty$. 

The Einstein-Hilbert action with a bulk cosmological constant and
a brane tension in the RSII model does not allow a self-tuning
solution. Addition of the Gauss-Bonnet term~\cite{kkl1} does not
improve the situation, and also it does not help to regularize
a naked singularity in the self-tuning model with a bulk
scalar~\cite{kachru}. But addition of the three index antisymmetric
tensor field $A_{MNP}$ (and the field strenth $H_{MNPQ}$ and $H^2\equiv
H_{MNPQ}H^{MNPQ}$) allows a self-tuning solution~\cite{kklcc}.

In this paper, we discuss the RSII model with 
antisymmetric tensor field added. The case with $1/H^2$
term allowing the self-tuning solution is the main motivation for this
extensive study. $H_{MNPQ}$ has been considered before in connection 
with the cosmological constant problem~\cite{witten} and the possible
compactification of the seven internal space in the
11D supergravity~\cite{fr}. Even in the 5D Randall-Sundrum model 
$H_{MNPQ}$ is useful to separate the 4 dimensional space. The $1/H^2$ 
term looks strange, but the consideration of the energy-momentum tensor 
would require a nonvanishing $\langle H_{MNPQ}\rangle$, 
triggering the separation of 
the extra dimension from the 4 dimensional space.
Below the Planck scale, we consider that the action is an effective
theory. Above the Planck scale, we consider that the quantum gravity
effects may be very important, but at present the final form for
quantum gravity is not known yet. 

In this paper, {\it the self-tuning solution means that it does
not need a fine-tuning between the parameters in the action}, which
is a progress toward understanding the cosmological constant
problem. The self-tuning solutions are found from time independent
Einstein equations. However, the existence of the 
self-tuning solution alone does not solve
the cosmological constant problem completely. It is because if the
ansatz for a {\it time dependent metric} allows, for example, the de 
Sitter space solutions with the antisymmetric tensor field added then
choosing the flat space is simply choosing a boundary condition. 
However, the existence of the {\it self-tuning solution} and the 
probabilistic interpretation for the wave function of the universe
can provide a logical understanding of the vanishing cosmological 
constant even in this case~\cite{hawking}. 
Suppose that we start from a flat universe 
from the beginning {\it $\acute a$ la} the wormhole interpretation 
of the vanishing cosmological constant. But this interpretation alone 
may encounter a difficulty when the phase transitions such as the 
electroweak phase transition or the QCD phase transition add 
a nonvanishing cosmological constant at a later epoch. However, 
the existence of the self-tuning solution chooses the flat space 
solution out of numerous possibilities when the universe goes 
through these phase transitions. If there exists a self-tuning 
solution, then the wormhole interpretation chooses the vanishing 
cosmological constant even after these phase transitions.

We find that the $1/H^2$ term {\it does} always allow
de Sitter space solutions with localized gravity.
In the RS model, however, a {\it negative} brane tension ($\Lambda_1<0$)
does not allow a localized gravity in the de Sitter space~\cite{karch},
but allows only a nonlocalized gravity.
There are arguments excluding these nonlocalized gravity~\cite{local},
and in the RSII model for a negative brane tension one may
exclude the de Sitter space solutions. However, the nonlocalized
gravity cannot become a strong argument for the vanishing cosmological
constant. It simply means that the de Sitter space solution with
nonlocalized gravity cannot materialize to our universe. At some
point, we may invoke an anthropic principle or turn to a probabilistic
interpretation. Namely, as long as there exist solutions for nonzero
cosmological constants whether there results a localized gravity or not, 
we need a probabilistic interpretaion. For a probabilistic 
interpretation, the existence of the self-tuning solution is
crucial in choosing the flat universe.

In Sec. II, we present the flat space solutions with $1/H^2$ term and
with $H^2$ term. It is shown that $1/H^2$ term allows the flat space
self-tuning solution but $H^2$ term allows at most one-fine-tuning 
solutions.  De Sitter space and anti de Sitter space solutions 
are also commented. 
In Sec. III, it is shown that the $1/H^2$ 
term allows the anti de Sitter and de Sitter space solutions. 
We discuss the horizons appearing in our solutions. We also discuss 
how the universe chooses the vanishing cosmological constant.
In Sec. IV, we present a time-dependent solution such that the 
4D space time remains {\it flat} when the brane tension shift 
instantaneously to another value. 
Sec. V is a conclusion.

\section{The static solutions}

The five dimensional space is composed of the bulk and
a 3-brane located at $y=0$ where $y$ is the fifth coordinate. 
We assume that matter fields
live in the brane. For studying the gravity sector, we
include the three index antisymmetric tensor field $A_{MNP}$ whose 
field strength is denoted as $H_{MNPQ}$, where $M,N,\cdots =0,1,2,3,
5(\equiv y)$. We find that there exist solutions for different bulk
cosmological constants at $y<0$ and $y>0$. But for simplicity
of the discussion, we will introduce a $Z_2$ symmetry so that the
bulk cosmological constant is universal. In this section, we summarize
the self-tuning solution ~\cite{kklcc} with the time independent
metric. But for comparison we briefly comment the time dependent 
metric, i.e. the de Sitter space and
anti de Sitter space solutions with the $1/H^2$ term. We also 
present one-fine-tuning solutions for $H^2$ term with time independent
metric, and compare with the other known tuning 
solutions~\cite{kachru,tamvakis}.

\subsection{A self-tuning solution of the cosmological constant with $1/H^2$}

A self-tuning solution exists for the following action,
\begin{equation}
S=\int d^4x\int dy \sqrt{-g}\left(\frac{1}{2}R
+\frac{2\cdot 4!}{H_{MNPQ}H^{MNPQ}}-\Lambda_b+{\cal L}_m\delta(y)
\right)\label{kklaction}
\end{equation}
which will be called the KKL model~\cite{kklcc}.
Here we set the fundamental mass parameter
$M$ as 1 and we will recover the mass $M$ wherever it is explicitly 
needed. We assume a $Z_2$ symmetry of the warp factor solution, 
$\beta(-y)=\beta(y)$. The sign of the $1/H^2$ term is chosen such
that at the vacuum the propagating field $A_{MNP}$ has a standard
kinetic energy term. 

The action contains the $1/H^2$ term which does not make sense if
$H^2$ does not develop a vacuum expectation value. 
Since the cosmological constant problem is at the bottom of most
cosmological application of particle dynamics, it is worthwhile to
study any solution to the cosmological constant problem. We note
that this problem has led to so many interesting but unfamiliar 
ideas~\cite{witten,baum,hawking,others}.  
Therefore, any new idea in the possible 
interpretation of the cosmological constant problem is acceptable at 
this stage. In fact, we found a very nice solution with the above 
action and hence we propose the action (\ref{kklaction}) as the 
fundamental one in gravity. Being a part of gravity, we do not
worry about the renormalizability at this stage.  

\vskip 0.5cm

\underline{Flat space solution} \hspace{0.2cm}
The ansatz for the metric is taken as
\begin{equation}
ds^2=\beta^2(y)\eta_{\mu\nu}dx^\mu dx^\nu+dy^2 \label{metric}
\end{equation}
where $(\eta_{\mu\nu})={\rm diag.}(-1,+1,+1,+1)$. 
Then Einstein tensors are,
\begin{eqnarray}
G_{\mu\nu}&=&g_{\mu\nu}\left[3\left(\frac{\beta^\prime}{\beta}\right)^2
+3\left(\frac{\beta^{\prime\prime}}{\beta}\right)\right],\nonumber\\
G_{55}&=&6\left(\frac{\beta^\prime}{\beta}\right)^2.
\end{eqnarray}
where prime denotes differentiation with respect to $y$.
With the brane tension $\Lambda_1$ at the $y=0$ brane and the bulk 
cosmological constant $\Lambda_b$, the energy momentum tensors are 
\begin{eqnarray}
T_{MN}= -g_{MN}\Lambda_b-g_{\mu\nu}\delta_M^\mu\delta_N^\nu
\Lambda_1 \delta(y)+4\cdot 4!\left(\frac{4}{H^4}H_{MPQR}H_N\,^{PQR}
+\frac{1}{2}g_{MN}\frac{1}{H^2}\right).\label{emten}
\end{eqnarray}

The specific form for $H^2\equiv H_{MNPQ}
H^{MNPQ}$ in Eq.~(\ref{kklaction}) 
makes sense only if $H^2$ develops a vacuum expectation
value at the order of the fundamental mass scale. Because of the
gauge invariant four index $H_{MNPQ}$, four space-time is 
singled out from the five dimensions~\cite{fr}. 
The four form field is denoted as $H_{\mu\nu\rho\sigma}$,
\begin{equation}
H_{\mu\nu\rho\sigma}=\sqrt{-g}\frac{\epsilon_{\mu\nu\rho\sigma}}{n(y)}
\label{g4}
\end{equation}
where $\mu,\cdots$ run over the Minkowski indices 0, 1, 2, and 3. With
the above ansatz, the field equation for the four form field is
satisfied,
\begin{equation}
\partial_M\left(\sqrt{-g}\frac{H^{MNPQ}}{H^4}\right)=0.
\end{equation}

There exists a solution for 
$\Lambda_b < 0$.  The two relevant Einstein equations are the (55) 
and ($\mu\mu$) components,
\begin{eqnarray}
6\left(\frac{\beta^\prime}{\beta}\right)^2&=&-\Lambda_b
- \frac{\beta^8}{A}\label{eqnbulk}\\
3\left(\frac{\beta^\prime}{\beta}\right)^2+3\left(
\frac{\beta^{\prime\prime}}{\beta}\right)&=&
-\Lambda_b-\Lambda_1\delta(y)-3\frac{\beta^8}{A}\label{eqnbulk1}
\end{eqnarray}
where $A$ is a positive constant in view of Eq.~(\ref{g4}).
It is easy to check that Eq.~(\ref{eqnbulk1}) in the bulk is obtained 
from Eq.~(\ref{eqnbulk}) for any $\Lambda_b,\Lambda_1,$ and $A$. 
This property is of the specialty of the $H_{MNPQ}$ field. 
Near B1(the $y=0$ brane), the $\delta$ function must be
generated by the second drivative of $\beta$. The
$Z_2$ symmetry, $\beta(-y)=\beta(y)$, implies
$(d/dy)\beta(y)|_{0^+}=-(d/dy)\beta(y)|_{0^-}$. Thus,
\begin{equation}
\frac{d^2}{dy^2}\beta(|y|)=\frac{d^2}{dy^2}\beta(|y|)\Big|_{
y\ne 0}+2\delta(y)\frac{d}{d|y|}\beta(|y|).
\end{equation}
This $\delta$-function condition at B1 leads to a boundary condition
\begin{equation}
\frac{\beta^\prime}{\beta}\Big|_{y=0^+}\equiv -k_1,\label{bc1}
\end{equation}
where we define $k$'s in terms of the bulk cosmological constant
and the brane tension,
\begin{equation}
k\equiv\sqrt{-\frac{\Lambda_b}{6}},\ \ \ 
k_1\equiv \frac{\Lambda_1}{6}.\label{k}
\end{equation}
Let us find a solution for the bulk equation
Eq.~(\ref{eqnbulk}) with the boundary condition Eq.~(\ref{bc1}). 
We define $a$ in terms of $A$,
\begin{equation}
a=\sqrt{\frac{1}{6A}}.\label{aa}
\end{equation}

The solution of Eq.~(\ref{eqnbulk}) consistent with the $Z_2$ symmetry is
\begin{equation}
\beta(|y|)=\left(\frac{k}{a}\right)^{1/4} {[\cosh(4k|y|+c)]^{-1/4}},
\label{solution}
\end{equation}
where $c$ is an integration constant to be determined by the 
boundary condition Eq.~(\ref{bc1}). This solution, consistent
with Condition (i), is possible for any value of
the brane tension $\Lambda_1$. Note that $c$ can take any sign.
This solution gives a localized
gravity consistent with the above Condition (ii).
The boundary condition (\ref{bc1}) determines $c$ in terms of 
$\Lambda_b$ and $\Lambda_1$,
\begin{equation}
c=\tanh^{-1}\left(\frac{k_1}{k}\right)=\tanh^{-1}
\left(\frac{\Lambda_1}{\sqrt{-6\Lambda_b}}\right).\label{c}
\end{equation}
A schematic shape of $\beta(y)$ is shown in Fig. 1.
\begin{figure}[t]
\label{fig:flat}
\includegraphics[width=100mm]{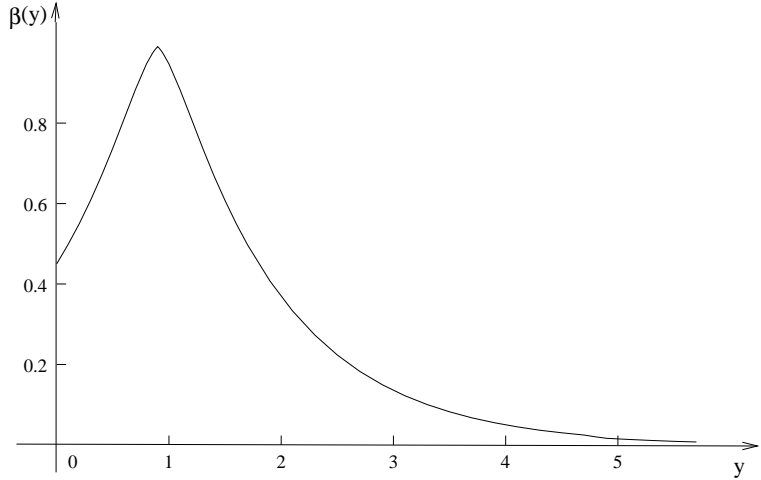}
\caption{$\beta(y)$ as a function of $y$ for the flat
space ansatz. It is plotted for $k=1$ and $a=1$.} 
\end{figure}

The effective 4D Planck mass is finite
$$
M_{P,\rm eff}^2=2M^3\left(\frac{k}{a}\right)^{1/2}\int_0^\infty
dy\frac{1}{\sqrt{\cosh(4ky+c)}}=\frac{M^3}{\sqrt{2ka}}F\left[\alpha,
\frac{1}{\sqrt{2}}\right]_0^\infty
$$ 
\begin{equation}
=\frac{M^3}{\sqrt{2ka}}
\int^1_{\sqrt{1-(\cosh(c))^{-1}}}\frac{dx}{\sqrt{(1-x^2)(1
-{1\over 2}x^2)}}.\label{planck}
\end{equation}
Here $F(\alpha,r)$ is the elliptic integral of the first kind and
\begin{equation}
\alpha=\sin^{-1}\sqrt{(\cosh (4ky+c)-1)/(\cosh (4ky+c))}. 
\end{equation}

Note that the Planck 
mass is given in terms of the integration constant $a$, or 
the integration constant is expressed in
terms of the fundamental mass $M$ and the 4D Planck mass $M_{P,\rm eff}$,
\begin{equation}
a=\left(\frac{M^3}{M_{P,\rm eff}^2\sqrt{2k}}F\left[\alpha,
\frac{1}{\sqrt{2}}\right]_0^\infty\right)^2.\label{bca}\label{A}
\end{equation}

\vskip 0.5cm

\underline{Curved space solution} \hspace{0.2cm}
The curved space solution is a time-dependent solution
which will be discussed in the KKL model of Sec. III.

\vskip 0.5cm

\underline{Localization of gravity and no tachyon} \hspace{0.2cm}
The perturbed metric is
\begin{eqnarray}
ds^2=(\beta^2\eta_{\mu\nu}+h_{\mu\nu})dx^\mu dx^\nu+dy^2\label{pmetric}
\end{eqnarray}
where we chose the Gaussian normal condition, $h_{5\mu}=h_{55}=0$.
With the transverse traceless gauge, $\partial^\mu h_{\mu\nu}=h^\mu_\mu=0$, 
and by a separation of variables such as 
$h_{\mu\nu}=\epsilon_{\mu\nu}e^{ipx}\psi(y)$ ($p^2=-m^2$),  
we obtain the following linearized equations without matter on the brane,
\begin{eqnarray}
\bigg[-\frac{1}{2}\beta^{-2}m^2-\frac{1}{2}\partial^2_y+2k^2-2k_1
\delta(y)-6a^2\beta^8\bigg]\psi(y)=0\label{lineq}
\end{eqnarray}
where 
\begin{eqnarray}
k_1&\equiv&\frac{\Lambda_1}{6}, \\
a&\equiv&\frac{1}{\sqrt{6A}}.
\end{eqnarray}
Here we make a change of variable by $z=\int^y\frac{dy}{\beta(y)}
\equiv\int^y du\,e^{-C(u)}$ and 
$\psi(y)=\beta^{1/2}\hat{\psi}(z)$ and then obtain 
the Schr$\ddot{\rm o}$dinger-like equation as follows,
\begin{eqnarray}
\bigg[-\frac{\partial^2}{\partial z^2}+V(z)\bigg]\hat{\psi}(z)=m^2\hat{\psi}(z) 
\end{eqnarray}
where 
\begin{eqnarray}
V(z)&=&\frac{15}{4}k^2\beta^2-3k_1\beta\delta(z)
-\frac{39}{4}a^2\beta^{10} \\
&=&\frac{9}{4}\bigg(\frac{\partial C}{\partial z}\bigg)^2
+\frac{3}{2}\bigg(\frac{\partial^2 C}{\partial z^2}\bigg).
\end{eqnarray}
Therefore, we can see that $m^2\geq 0$, i.e., there is no tachyon state near 
the background, by regarding the above equation as
a supersymmetric quantum mechanics~\cite{tamvakis},
\begin{eqnarray}
Q^\dagger Q\hat{\psi}(z)\equiv\bigg(\partial_z+\frac{3}{2}\frac{\partial C}
{\partial z}\bigg)
\bigg(-\partial_z+\frac{3}{2}\frac{\partial C}{\partial z}\bigg)\hat{\psi}(z)
=m^2\hat{\psi}(z).
\end{eqnarray}
Moreover, there appears the localization of gravity on the brane as expected 
from the finite 4D Planck mass, 
due to one bound state of massless graviton, $\hat{\psi}_0(z)\sim 
e^{3C/2}=\beta^{3/2}$, that is, $\psi_0(y)\sim \beta^2$.  
Note that the zero mode solution $\psi_0(y)$ automatically satisfies 
the boundary condition $(\partial_y+2k_1)\psi|_{y=0+}=0$ from Eq.~(\ref{lineq}).

\subsection{One-fine-tuning solutions of the cosmological constant
with $H^2$ term}

In this section, we obtain just the flat space solution
for the $H^2$ term in the action,
\begin{equation} \label{action1}
S=\int d^4x\int dy \sqrt{-g}\left(\frac{M^3}{2}R-\frac{M}{2\cdot 4!}
H_{MNPQ}H^{MNPQ}-\Lambda_b+\sum_{i} {\cal L}_m^{(i)}\delta(y-y_i)
\right).  
\end{equation}
The ansatze for the metric and the four form field are taken also as given 
in Eq.~(\ref{metric}) and Eq.~(\ref{g4}), respectively.  
Thus the field equation for $A_{MNP}$ is trivially satisfied again, 
\begin{equation}
\partial_{M}\bigg[\sqrt{-g}H^{MNPQ}\bigg]=0.
\end{equation}  
The two relevant Einstein equations are the (55) and ($\mu\mu$) components,
\begin{eqnarray}
6\left(\frac{\beta^\prime}{\beta}\right)^2&=&-\Lambda_b
+\frac{A}{\beta^8}\label{eqnb}\\
3\left(\frac{\beta^\prime}{\beta}\right)^2+3\left(
\frac{\beta^{\prime\prime}}{\beta}\right)&=&
-\Lambda_b-\Lambda_1\delta(y)-\Lambda_2\delta(y-y_c)
-\frac{A}{\beta^8}\label{eqnb1}
\end{eqnarray}
where $A/\beta^8\equiv 1/2n^2$ expressed in terms of a `positive' constant $A$.  

The solutions of Eq.~(\ref{eqnb}) and (\ref{eqnb1}) are 
\begin{eqnarray}
{\rm for}~~~\Lambda_b<0~~:~~\label{nega}
\beta(|y|)&=&\left(\frac{a}{k}\right)^{1/4} {[\sinh(| 4k|y|+c|)]^{1/4}} \\
{\rm for}~~~\Lambda_b>0~~:~~\label{posi}
\beta(|y|)&=&\left(\frac{a}{k}\right)^{1/4} {[\sin(| 4k|y|+c|)]^{1/4}} \\
{\rm for}~~~\Lambda_b=0~~:~~\label{zero}
\beta(|y|)&=&\left(| 4a|y|+c|\right)^{1/4},
\end{eqnarray} 
where the $a$ is defined in terms of $A$, 
\begin{equation}
a\equiv \sqrt{\frac{A}{6}}.\label{a}
\end{equation}
We note that for a positive $c$ in Eqs.~(\ref{nega})--
({\ref{zero}) $\beta$'s do not give localized graviton solutions near the 
brane B1, and for a negative $c$ $\beta$'s have naked 
singularities at $|y|=-c/4k$ or $-c/4a$.  
Therefore, to get the effective four dimensional gravity or to avoid the 
sigularities in the bulk, it is indispensable to cut the extra dimension 
such that it has a finite length size by introducing another brane,
say B2. Thus, we need at least two branes and the situation is similar 
to that of the RSI except for the 4 form field contributions.   
Since the extra dimension is finite, the effective four dimensional 
Planck mass $M_P\equiv M^3\int dy \beta^2$ is also finite.  
If we introduce two branes, 
we should satisfy the boundary conditions at the two branes, 
consistently with the $S^1/Z_2$ orbifold symmetry, 
\begin{equation}
\frac{d^2}{dy^2}\beta(|y|)=\frac{d^2}{dy^2}\beta(|y|)\Big|_{
y\ne 0}+2\left(\delta(y)-\delta(y-y_c)\right)\frac{d}{d|y|}\beta(|y|).
\end{equation}
Then the boundary conditions for the above three cases are 
\begin{eqnarray}
{\rm for}~~~\Lambda_b<0~~:~~
c&=&{\rm coth}^{-1}\left(\frac{k_1}{k}\right)
=4ky_c-{\rm coth}^{-1}\left(\frac{k_2}{k}\right) \label{onetune2}\\
{\rm for}~~~\Lambda_b>0~~:~~
c&=&{\rm cot}^{-1}\left(\frac{k_1}{k}\right)
=4ky_c-{\rm cot}^{-1}\left(\frac{k_2}{k}\right) \label{onetune1}\\
{\rm for}~~~\Lambda_b=0~~:~~
c&=&\frac{a}{k_1}=a\left(4y_c-\frac{1}{k_2}\right),\label{onetune} 
\end{eqnarray}    
where $k_2$ is defined in terms of the brane tension $\Lambda_2$
at B2, 
\begin{equation}
k_2\equiv \frac{\Lambda_2}{6}. 
\end{equation} 
We note that in the case of the $H^2$ term (not $1/H^2$) in the action, 
the one-fine-tuning relations between $k_1$ and $k_2$ appear always,
e.g. the relations (\ref{onetune2},\ref{onetune1},\ref{onetune}), while 
in the case of RSI model, the two-fine-tuning relations $k=k_1=-k_2$ 
were inevitable. 

If we introduce both $1/H^2$ and $H^2$ in the action,
there does not exist a flat space self-tuning solution. In this case,
the derivative of the warp factor satisfies
\begin{equation}
\beta^\prime= \pm\sqrt{\bar\Lambda +\frac{\Lambda_b \beta^2}{6}
-\frac{\beta^8}{6A}+\frac{A}{6\beta^8}}.\label{HH}
\end{equation}
A necessary condition for the existence of a 
flat space solution is $\beta^\prime\rightarrow 0$
and $\beta\rightarrow 0$ as $y\rightarrow \infty$. Certainly,
Eq.~(\ref{HH}) does not satisfy this necessary condition.
On the other hand, a necessary condition for the de Sitter
space regular horizon is that $\beta^\prime={\rm finite}$ as 
$\beta\rightarrow 0$. Also, this condition is not met in
Eq.~(\ref{HH}) and hence there does not exist de Sitter
space regular horizon. 

\subsection{Comparison with other tuning solutions}

There appeared several self-tuning solutions since early eighties
\cite{witten,hawking,kachru,tamvakis}. The early stage scenario
\cite{witten,hawking} used field strength of
three index antisymmetric tensor field 
$H_{\mu\nu\rho\sigma}$ to introduce an
integration constant. The vacuum value of $\langle H_{\mu\nu\rho\sigma}
\rangle= \epsilon_{\mu\nu\rho\sigma}c$ 
introduces an integration constant $c$
which contributes to the cosmological constant $\propto c^2$. 
Therefore, there exists a value of $c$ such that the effective
cosmological constant vanishes for a range of bare cosmological 
constant. Once $c$ is determined to give the zero effective 
cosmological constant, $c$ cannot change since it is a constant.
In this sense, the four dimensional example is not a working model.
Thus, the self-tuning solution needed a dynamical field to
propagate. The three index antisymmetric field is not a dynamical
field in 4D spacetime.

Introduction of the extra dimension opened a new game in the
self-tuning solutions of the cosmological 
constants~\cite{kachru,kklcc,tamvakis,tye,cline,davis}. Here, we briefly
comments on the key points of these solutions.

The California self-tuning solution~\cite{kachru,nilles} must use
the RSI model set up, i.e. introduce two branes B1 and B2,
and introduce a specific form for the
potential of a bulk scalar field $\phi$ coupling to the brane
matter with a desired form.
But the model has a naked singularity and to cure the problem,
one has to introduce a brane at this singular point. Then,
one needs a condition at the new brane and leads to one fine-tuning
there~\cite{nilles}. 
If one satisfies this one-fine-tuning condition between
parameters in the Lagrangian, there always
exists a flat space solution. Therefore, the solution is
not a self-tuning solution originally anticipated.
One interesting or (disastrous to some) point of this model is that it
does not allow the de Sitter or anti de Sitter space solutions.
In this sense, it is an improvement over the original proposal
\cite{witten,hawking}
for the self-tuning solutions. However, it may be difficult to
obtain a period of inflation since the de Sitter space exponential
expansion is not possible. One should see whether the bulk energy
momentum tensor satisfies $\rho=-p=-p_5/2$ to have the exponential
expansion of the effective 4D space~\cite{kyae}. However, at
present it is not known what matter satisfies this kind of 
equation of state.

There appeared another interesting self-tuning 
solution~\cite{tamvakis,earlier}, which does not introduce any brane.
This model assumes a specific form of the 5D scalar potential  
multiplied to the matter Lagrangian in 5D. Even though the metric
gives a localized gravity, the matter fields propagate in the full 5D
space since the potential tends to a constant value as
$y\rightarrow\infty$. It is not certain how we are forbidden to 
realize the extra dimension in this model.

Then, there are proposals that the de Sitter space solution is
phenomenologically acceptable as far as the curvature at present
is sufficiently small~\cite{hbk,tye,cline,davis}. In a sense,
it also tries to accomodate the current small vacuum energy
\cite{perl} with the de Sitter space solutions. [On the other hand,
note that the quintessence idea is based on the solution 
of the cosmological constant problem~\cite{quint}.]
The way the proposals make the vacuum energy small is to separate
the distance between branes sufficiently large since the vacuum 
energy is exponentially decreasing with the separating distance.
The one fine-tuning at B1 is met but the second fine tuning at
B2 is not satisfied and allows the de Sitter space solutions.
To have the vacuum energy decreased down to the current
energy density, one needs that the separation distance increases
as $t$ increases. Namely, the horizon point at the $y$ axis
is required to increase. Then it is possible to relate the
current vacuum energy with the current mass energy. 
In Ref.~\cite{davis}, the relation $\Omega_\Lambda=(5/2)\Omega_m$
is obtained for $\Omega_{total}=1$.

The model presented in \cite{kklcc} allows a self-tuning flat
space solution, self-tuning de Sitter and anti de Sitter space
solutions. In general it is possible to introduce a period of
inflation. The transition from one flat space to another flat
space can arise following the solutions of the Einstein and
field equations, as we discuss in the subsequent sections.

\section{De Sitter space solutions}

We find that there exist de Sitter space solutions in our model with a
simple time dependence. First, let us briefly discuss the de Sitter
space solution~\cite{karch} in the RS II model and present
de Sitter space solutions in our model.

\subsection{The RS-II model}

Let us consider the following metric ansatz for the $dS_{4}$ brane 
in the RS II model:
\begin{eqnarray}
&ds^2=\beta^2(y)[-dt^2+e^{2\sqrt{\bar{\Lambda}}t}\delta_{ij}dx^i dx^j]+dy^2,
\nonumber \\
&=\beta^2(y)\hat{g}_{\mu\nu}dx^\mu dx^\nu+dy^2
\end{eqnarray}
from which the 4D Ricci tensor is given as 
$R^{(4)}_{\mu\nu}=3\bar{\Lambda}\hat{g}_{\mu\nu}$. 
Then, the warp factor $\beta(y)$ of the $dS_{4}$ brane solution is given by
\begin{eqnarray}
\beta(y)&=&\frac{\sqrt{\bar{\Lambda}}}{k}\sinh[k(y_m-y)],
\nonumber \\ 
y_m&=&\frac{1}{k}\coth^{-1}\bigg(\frac{k_1}{k}\bigg)
\end{eqnarray}
where the integration constant $y_m$ is determined from the boundary condition 
at the brane and  
\begin{eqnarray}
k\equiv\sqrt{-\frac{\Lambda_b}{6}}, \,\,\, k_1\equiv \frac{\Lambda_1}{6}. 
\end{eqnarray}
For a positive tension brane, we get a positive value of $y_m$, 
which gives rise to an event horizon at the finite proper distance away 
from the brane.  Therefore, the region beyond the horizon can not be 
causally connected to the region where the brane resides. This bulk 
horizon resembles the $dS_4$ horizon on the brane.
And, the $dS_4$ solution is regular along the bulk
because the curvature tensors become finite at the bulk horizon. To say about 
whether the $dS_4$ solution could describe a dynamical compactification of the 
extra dimension with the region beyond the horizon cut off, we should check 
its consistency from the 4D effective cosmological constant by integrating out 
the fifth dimension. 
The 4D effective action is given by 
\begin{eqnarray}
S_{4,eff}&\simeq&\int d^4 x\sqrt{-g^{(4)}}\int^{y_m}_{-y_m}dy\,
\beta^4\bigg[\frac{1}{2} R^{(5)}-\Lambda_b-\Lambda_1\delta(y)\bigg],
\nonumber\\
&\simeq&\int d^4 x\sqrt{-g^{(4)}}\bigg(\frac{M^2_{P,eff}}{2}R^{(4)}
-6\Lambda_{eff}\bigg)
\end{eqnarray}   
where $g^{(4)}$ and $R^{(4)}$ are given from $g_{\mu\nu}=\hat{g}_{\mu\nu}
+h_{\mu\nu}$ in the zero mode 
expansion and the 4D effective Planck mass and the 4D effective cosmological
constant are determined as follows, 
\begin{eqnarray}
M^2_{P,eff}=\frac{\bar\Lambda}{k^3}\bigg(-ky_m+\frac{1}{2}
\sinh(2ky_m)\bigg)>0
\end{eqnarray}
and
\begin{eqnarray}
\Lambda_{eff}&=&\frac{1}{3}\int^{y_m}_{-y_m}dy\,\beta^4
\bigg(4\frac{\beta^{\prime\prime}}{\beta}+6\bigg(\frac{\beta^\prime}{\beta}
\bigg)^2+\Lambda_b+\Lambda_1\delta(y)\bigg), \nonumber\\
&=&\frac{\bar{\Lambda}^2}{k^3}\bigg(-ky_m+\frac{1}{2}\sinh(2ky_m)\bigg)>0.
\end{eqnarray} 
Then, when we compare the 4D Einstein equations of motion derived from 
the above 4D effective action with those satisfied by the $dS_4$ brane 
solution, we can show that the ratio $\Lambda_{eff}/M^2_{P,eff}$ is 
exactly equal to the 4D cosmological constant $\bar{\Lambda}$ of the brane 
solution. Therefore, the $dS_4$ brane solution can be regarded to 
reproduce the 4D effective de Sitter spacetime with the extra dimension 
dynamically compactified up to the horizon distance in the bulk. 

It is easy to observe that $\Lambda_1<0$ does not allow a horizon,
which implies a nonlocalized gravity.

\subsection{The KKL model}

The background metric is
\begin{eqnarray}
ds^2=\beta^2(y)\bar{g}_{\mu\nu}dx^\mu dx^\nu+dy^2\label{backgr}
\end{eqnarray}
where the metric corresponds to one of the 4D maximally symmetric
$\bar \Lambda\ne 0$ spaces: 
$$
\bar{g}_{\mu\nu}={\rm diag}(-1,e^{2\sqrt{\bar{\Lambda}}t},
e^{2\sqrt{\bar{\Lambda}}t},e^{2\sqrt{\bar{\Lambda}}t})\ ,\ \ (dS_4\
\rm  background,\  \bar{\Lambda}>0),
$$
for de Sitter space, and
$$
\bar{g}_{\mu\nu}=
{\rm diag}(-e^{2\sqrt{-\bar{\Lambda}}x_3},
e^{2\sqrt{-\bar{\Lambda}}x_3},e^{2\sqrt{-\bar{\Lambda}}x_3},1)\ ,
\ \ (AdS_4\ \rm background, \bar{\Lambda}<0), 
$$
for anti-deSitter space.

The 4D Ricci tensor is given as 
$\bar{R}_{\mu\nu}=3\bar{\Lambda}\bar{g}_{\mu\nu}$.
And, components of the five dimensional Einstein tensor are
\begin{eqnarray}
G_{\mu\nu}&=&\bigg[3\bigg(\frac{\beta^{\prime\prime}}{\beta}\bigg)
+3\bigg(\frac{\beta^\prime}{\beta}\bigg)^2-3\bar{\Lambda}\beta^{-2}\bigg]
\beta^2\bar{g}_{\mu\nu}, \\
G_{55}&=&6\bigg(\frac{\beta^\prime}{\beta}\bigg)^2-6\bar{\Lambda}\beta^{-2}.
\end{eqnarray} 
where $\bar\Lambda>0$ for $dS_4$ and $\bar\Lambda<0$ for $AdS_4$
background. The $\bar\Lambda$ term arises from the time derivatives 
of the metric. 
Then, the (55) and ($\mu\nu$) components of the Einstein's equations 
with $1/H^2$ follows,
\begin{eqnarray}
6\bigg(\frac{\beta^\prime}{\beta}\bigg)^2-6\bar{\Lambda}\beta^{-2}
&=&-\Lambda_{b}-\frac{\beta^8}{A}, \\
3\bigg(\frac{\beta^\prime}{\beta}\bigg)^2+3\bigg(\frac{\beta^{\prime\prime}}
{\beta}\bigg)-3\bar{\Lambda}\beta^{-2}&=&-\Lambda_b-\Lambda_1\delta(y)
-3\frac{\beta^8}{A}.
\end{eqnarray}
We can obtain bulk solutions by solving the (55) component. 

We can consider two cases: $\beta=0$ at $y=y_m$=(finite) or 
$\beta\rightarrow 0$ as $y\rightarrow\infty$. Otherwise, the
4D Planck mass diverges and localizable gravity does not follow.
The case $\beta=0$ at a finite $y$ arises in the deSitter space 
solution.

\begin{figure}[t]
\label{fig:ads4}
\includegraphics[width=100mm]{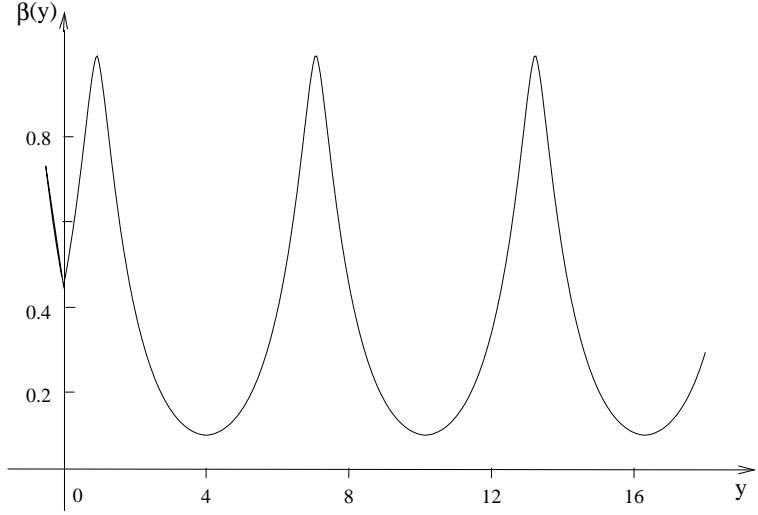}
\caption{$\beta(y)$ as a function of $y$ for the anti de Sitter
space solution. It is plotted for $\bar k^2=-0.01$, $k=1$
and $a=1$, where $\bar k^2=\bar\Lambda$.} 
\end{figure}

For the anti de Sitter space, let us 
consider the asymptotic behavior of the warp factor $\beta(y)$ 
as $y\rightarrow \infty$. With $z=\frac{1}{y}$, we can rewrite the (55) 
component as,
\begin{eqnarray}
\bigg(\frac{d\beta}{dz}\bigg)^2=\frac{\bar{\Lambda}}{z^4}
+\frac{k^2\beta^2}{z^4}-\frac{a^2\beta^{10}}{z^4},
\end{eqnarray} 
where 
\begin{eqnarray}  
k^2\equiv-\frac{\Lambda_b}{6}, \,\,\, a^2\equiv\frac{1}{6A}.  
\end{eqnarray}
To get the finite 4D Planck mass from the non-compact extra dimension or 
have the localized gravity, $\beta$ should be zero as $z\rightarrow 0$. 
However, the 4D cosmological constant term should be divergent as 
$z\rightarrow 0$ even if the last two terms are set to be zero. 
Therefore, there does not appear the localized gravity for the AdS 
background with non-zero effective 4D cosmological constants.

\vskip 0.5cm

In the KKL model~\cite{kklcc} the de Sitter space metric 
ansatz gives $\beta^\prime$ as
\begin{equation}
\beta^\prime=\sqrt{\bar k^2+k^2\beta^2-a^2\beta^{10}}\label{betaprime}
\end{equation}
where
\begin{equation}
\bar k=\sqrt{\frac{\bar\Lambda}{6}},\ \ k=\sqrt{\frac{-\Lambda_b}{6}}.
\end{equation}
The boundary condition at $y=0$ relates the brane tension ($k_1$)
and the bulk cosmological constant,
\begin{equation}
k_1=\sqrt{k^2-a^2\beta^8(0)+\frac{\bar\Lambda}{\beta(0)^2}}\label{dsbc}
\end{equation}
where $\beta(0)$ is a function of an integration constant $c$.
The above relation (\ref{dsbc}) is valid for both the de Sitter and anti
de Sitter space solutions.

Consider the half space $y>0$. If we choose the minus sign
in front of the square root of Eq.~(\ref{betaprime}), there always
exist $y_m$ as in the RSII model.

Even if we choose the plus sign in front of the square root of 
Eq.~(\ref{betaprime}) near $y=0^+$, there exist horizons. This is 
because there can exist a point where $\beta^\prime$ vanishes for a
sufficiently large $\beta$ in the region where $\beta$ increases. For this
not to be realized, $\beta$ should tend to an asymptotic limit
as $y\rightarrow\infty$. If it goes to an asymptotic limit, the
Planck mass becomes infinite and there does not result a localized
gravity. Therefore, we restrict to the case for $\beta$ turning around
and coming down, in which case there results a horizon which is
determined by the point where $\beta=0$ at $y=y_m$.
Between $y=0$ and $y=y_m$, there can ($\Lambda_1<0$)
or cannot ($\Lambda_1>0$) exist a point where $\beta^\prime=0$. 
Even if there exists
such a point as in the above example, 
the integral $\int d\beta/\sqrt{\bar k^2+k^2\beta^2
-a^2\beta^{10}}$ always converges, and we obtain a localized gravity
with the horizon at $y=y_m$. 
Therefore, there always exist de Sitter space solutions, which have the
periodic form as shown in Fig. 3. 
Similarly, there exist anti de Sitter space solutions, which
have the periodic form as shown in Fig. 2. But the anti de Sitter 
space solution does not lead to a localized gravity. 
Even if we allow the anti de Sitter space
solutions, neglecting the unphysical unlocalized gravity, the
probability to choose the anti de Sitter space is exponentially small
~\cite{hawking}. 

\begin{figure}[t]
\label{fig:ds4}
\includegraphics[width=100mm]{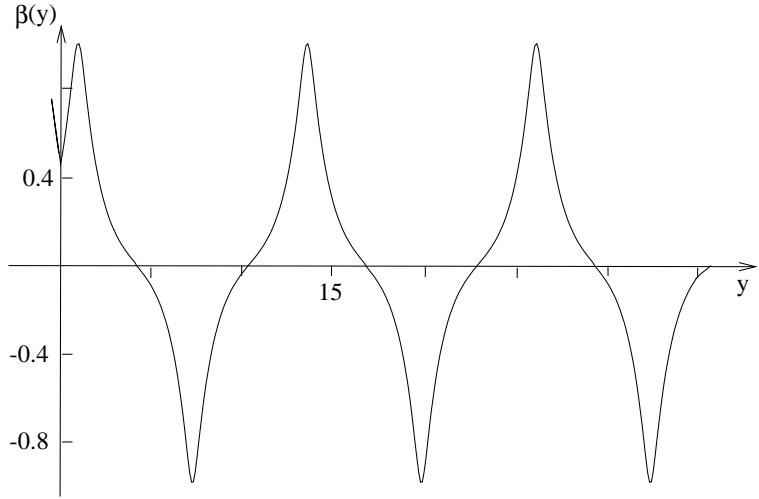}
\caption{$\beta(y)$ as a function of $y$ for the de Sitter
space solution. It is plotted for $\bar k^2=0.01$, $k=1$
and $a=1$, where $\bar k^2=\bar\Lambda$.} 
\end{figure}

The curved space solutions in the KKL model have the effective bulk 
cosmological constant more negative compared to the RSII case.
Therefore, the bound from the weaker energy 
condition on the background matter supporting those spacetimes is completely 
satisfied. On the other hand the weaker energy condition
is saturated in the RSII case~\cite{karch}.
The weaker energy condition reads
$T_{MN}\xi^M\xi^N\ge 0$ for a null vector $\xi^M$, which corresponds to 
$(\beta^\prime/\beta)^\prime\le -\bar{\Lambda}\beta^{-2}$ 
for the background metric Eq.~(\ref{backgr}). In the KKL case, the weaker 
energy condition is always satisfied without the possibility of
saturation,
\begin{eqnarray}
\bigg(\frac{\beta^\prime}{\beta}\bigg)^\prime=-\bar{\Lambda}\beta^{-2}
-4a^2\beta^8< -\bar{\Lambda}\beta^{-2}.
\end{eqnarray}

Existence of de Sitter space solutions in the KKL model 
allows possible cosmological constants in 4D space as {\it nonnegative}. 
Hawking's probabilistic interpretation of the cosmological
constant chooses the flat space~\cite{hawking}. However, we need
a qualification in this statement. Existence of nonlocalizable
de Sitter space solution can give a large probability. The
Euclidean action can be written schematically as 
\begin{equation}
S_E\propto \int d^4x \left(M_P^2 \frac{R_4}{2}+\cdots\right),
\end{equation}
after the $y$ integration. The de Sitter space solutions have
horizons as shown in Fig. 3. If we integrate over the whole
region of the $y$ space $M_P$ diverges. But the Planck mass
must be from the integration only up to the horizon connected
to $y=0$ brane which we call the first universe. Then $M_P$ is finite
in the first universe. However, if we neglect the effect of
the brane, the probability to
go the the second universe (the universe between the next two 
horizons) is of the same order as the probability to go to the first
de Sitter space universe, because the horizen points appear 
periodically. Summing up the probabilities to go to either of
these de Sitter universes, we would obtain an infinity. Nevertheless,
the second universe does not contain the brane, and we have to
exclude this possibility of transition to the universes not
containing the brane. Therefore, $M_P$ is considered to be finite,
and the probability to stay in the flat universe is 
maximum~\cite{hawking}. 

\section{Time-dependent solution of the self-tuning model}

The existence of the self-tuning solution will be a great leap
toward the understanding of the vanishing cosmological constant
~\cite{hawking,witten}.  As shown in the previous section,
the action given in Ref.~\cite{kklcc}
also allows de Sitter and anti de Sitter space solutions.

Note that if there does not exist any de Sitter space 
and anti de Sitter space solutions but there exist
flat space self-tuning solutions then the
cosmological constant may become {\it automatically} zero
for a finite range of parameters in the Lagrangian.
In this sense, our self-tuning solution does {\it not}
lead to a vanishing cosmological constant {\it automatically} but 
the cosmological constant is {\it probably zero} {\it $\acute a$ la} 
Hawking~\cite{hawking}. However, the automatic solutions have
a difficulty in implementing inflationary period which is
needed for our sufficiently homogeneous and isotropic 
universe~\cite{guth}.

It is anticipated that the spontaneous symmetry breaking at
the brane proceeds at the electroweak phase transition 
and the QCD phase transition. Then the effective potential
energy will have a time dependence of the form, 
$\Lambda_1(t)\delta(y)$. According to the shift of the potential
energy at the brane, the solution of the field and Einstein
equations will have a time-dependence. Certainly, there will
exist time-dependent solutions, and we will be
interested in solutions {\it $\acute a$ la} Hawking~\cite{hawking}, 
changing from one flat space solution with an integration constant 
$c_1$ corresponding to the brane cosmological constant $\Lambda_{old}$ 
to another integration constant $c_2$ corresponding to the brane 
cosmological constant $\Lambda_{new}$. This situation corresponds 
to a time dependent energy momentum tensor. This transition from 
a flat space solution to another flat space solution is through
satisfying the field equations and is different
from Witten's sudden choice of a flat space solution~\cite{witten}.
There can be time dependent solutions from a flat space to de Sitter
or anti de Sitter spaces, but the probability for these transitions
is exponentially small compared to a flat to flat 
transition~\cite{hawking}.

To study the time-dependence the metric is taken as
\begin{equation}
ds^2=-n^2(t,y)dt^2+a^2(t,y)\eta_{ij}dx^i dx^j
+b^2(t,y)dy^2.\label{tmetric}
\end{equation}
The Einstein tensors are,
$$
G_{00}=g_{00}\left\{-\frac{3}{n^2}\frac{\dot a}{a}\left(\frac{\dot a}{a}
+\frac{\dot b}{b}\right)+\frac{3}{b^2}\left(\frac{a^{\prime\prime}}{a}
+\frac{a^\prime}{a}\left(\frac{a^\prime}{a}-\frac{b^\prime}{b}
\right)\right)\right\}
$$
$$
G_{ii}=g_{ii}\Big\{-\frac{1}{n^2}\left[2\frac{\ddot a}{a}
+\frac{\ddot b}{b}-\frac{\dot a}{a}\left(2\frac{\dot n}{n}
-\frac{\dot a}{a}\right)-\frac{\dot b}{b}\left(\frac{\dot n}{n}
-2\frac{\dot a}{a}\right)\right]
$$
\begin{equation}
+\frac{1}{b^2}\left[\frac{n^{\prime\prime}}{n}+2\frac{a^{\prime\prime}}{a}
+\frac{a^\prime}{a}\left(2\frac{n^\prime}{n}+\frac{a^\prime}{a}\right)
-\frac{b^\prime}{b}\left(\frac{n^\prime}{n}+2\frac{a^\prime}{a}\right)\right]
\Big\} \label{teinstein}
\end{equation}
$$
G_{55}=g_{55}\left\{-\frac{3}{n^2}\left(\frac{\ddot a}{a}
-\frac{\dot a}{a}\left(\frac{\dot n}{n}-\frac{\dot a}{a}\right) \right)
+\frac{3}{b^2}\frac{a^\prime}{a}\left(\frac{n^\prime}{n}
+\frac{a^\prime}{a}\right)\right\}
$$
$$
G_{05}=3\left(\frac{\dot a}{a}\frac{n^\prime}{n}+\frac{\dot b}{b}
\frac{a^\prime}{a}-\frac{\dot a^\prime}{a}\right) 
$$
where dot and prime denote differentiations with respect to $t$
and $y$, respectively.

With the brane tension $\Lambda_1$ at the $y=0$ brane and the bulk 
cosmological constant $\Lambda_b$, the energy momentum tensor is 
\begin{eqnarray}
T_{MN}= -g_{MN}\Lambda_b-g_{\mu\nu}\delta_M^\mu\delta_N^\nu
\Lambda_1 \delta(y)+4\cdot 4!\left(\frac{4}{H^4}H_{MPQR}H_N\,^{PQR}
+\frac{1}{2}g_{MN}\frac{1}{H^2}\right),\label{emtensor}
\end{eqnarray}
Considering the homogeneous 3D space, nonzero components of 
$H^{MNPQ}$ are, 
\begin{eqnarray}
&H^{\mu\nu\rho\sigma}=\frac{1}{\sqrt{-g}}\epsilon^{\mu\nu\rho\sigma 5} 
\partial_5\sigma\nonumber\\
&H^{5ijk}=\frac{1}{\sqrt{-g}}\epsilon^{5ijk0}\partial_0
\sigma.\label{hmnpq}
\end{eqnarray}
Then, we have
\begin{eqnarray}
&H^2=-4!(f^2-h^2),\ \ H_{0NPQ}H_0\,^{NPQ}=-3!g_{00}f^2,\nonumber\\
&H_{iNPQ}H_j\,^{NPQ}=-3!g_{ij}(f^2-h^2),\ \ H_{5NPQ}H_5\,^{NPQ}
=3!g_{55}h^2,\\
&H_{5NPQ}H_0\,^{NPQ}=3!bnfh,
\ \ H_{5NPQ}H_i\,^{NPQ}=0,\nonumber
\end{eqnarray}
where $f^2=\sigma^{\prime 2}/b^2$ and $h^2=\dot\sigma^2/n^2$.
The $T_{MN}$ appearing in the Einstein equations $G_{MN}=T_{MN}$ 
are
\begin{eqnarray}
&T_{00}=-g_{00}\left(\Lambda_b+\Lambda_1\delta(y)
+\frac{6}{f^2-h^2}+\frac{4h^2}{(f^2-h^2)^2}\right)\nonumber\\
&T_{ij}=-g_{ij}\left(\Lambda_b+\Lambda_1\delta(y)
+\frac{6}{f^2-h^2}\right)\nonumber\\
&T_{55}=-g_{55}\left(\Lambda_b+\frac{2}{f^2-h^2}
-\frac{4h^2}{(f^2-h^2)^2}\right)\label{tensorcomp}\\
&T_{05}=\frac{4bnfh}{(f^2-h^2)^2}.\nonumber
\end{eqnarray}

The field equation of the four-form field is
$$
\partial_M \left[\frac{\sqrt{-g}H^{MNOP}}{H^4}\right]=
\partial_0\left[\frac{\epsilon^{0ijk5}\sigma^\prime}{H^4}\right]
+\partial_5\left[\frac{\epsilon^{5ijk0}\dot\sigma}{H^4}\right]
$$
\begin{equation}
=\frac{\epsilon^{0ijk5}}{(4!)^2}\left\{
\partial_0\left[\frac{\sigma^\prime}{((\sigma^{\prime}/b)^2-(\dot 
\sigma/n)^2)^2}\right]
-\partial_5\left[\frac{\dot\sigma}{((\sigma^\prime/b)^2-(\dot 
\sigma/n)^2)^2}
\right]\right\}=0.\label{fieldeq}
\end{equation}

Note that $\sigma$ is given, in the static homogeneous 4D case considered
above, by 
\begin{equation}
\sigma(|y|)=\frac{ab}{4k^2}\sqrt{2A} \sinh (4kb|y|+c),    
\end{equation}
with $2/f^2=\beta^8/A$.

It is very difficult to find a general solution of the above Einstein
equations. However, we are interested in the existence of
the interpolating solution between two flat space solutions.
We anticipate that a flat space solution, viz. Eq.~(\ref{solution}), 
with $c_i=\tanh^{-1}(\Lambda_i/\sqrt{-6\Lambda_b})$ is given, probably
by the initial condition or Hawking's choice of the cosmological
constant. During a phase transition when $\Lambda_{old}$ changes to
$\Lambda_{new}$, we try to show the existence of an interpolating
solution connecting two flat space solutions with $c_1$ and $c_2$.

Let us consider the case that the brane tension 
$\Lambda_1=\Lambda_{old}$ changes to 
$\Lambda_1=\Lambda_{new}$ instantaneously due to a phase transition 
by {\it brane matter} at the brane. Then the boundary condition 
requires a time dependent $b$, 
\begin{equation}\label{b(t)}
b(t)=(b_{{\rm new}}-b_{{\rm old}})\theta(t-t_0)+b_{{\rm old}}, 
\end{equation}
which gives
\begin{equation}
\beta(|y|,t)=\left(\frac{k}{a}\right)^{1/4}
\left[{\rm cosh}\left(4kb(t)(|y|+y_1)\right)\right]^{-1/4},
~~~{\rm for} ~~t\ne t_0,  \label{beta(t)}  
\end{equation}
where $b_{old}, b_{new}$ and $y_1$ are constants. $c$ in 
Eq.~(\ref{solution}) is endowed with a time-dependence as given above.
The `$t$' dependence of $\beta$ generates non-zero time-derivatives 
of our metric in the Einstein equation. We suppose that the metric
dynamics gives rise to the {\it bulk matter} fluctuation near the vacuum, 
$T^{(m)~M}\,_{N}\equiv {\rm diag}\left(-\rho,~p,~p,~p,~p_5\right)$,  
\begin{eqnarray}
\rho&=&\frac{3}{\beta^2}\frac{\dot{\beta}}{\beta}
\left(\frac{\dot{\beta}}{\beta}
+\frac{\dot{b}}{b}\right)  \\
p&=&-\frac{1}{\beta^2}\left[2\frac{\ddot \beta}{\beta}
+\frac{\ddot b}{b}-\left(\frac{\dot \beta}{\beta}\right)^2  
+\frac{\dot b}{b}\frac{\dot \beta}{\beta}
\right]  \\
p_5&=&-\frac{3}{\beta^2}\left(\frac{\ddot \beta}{\beta}\right).
\end{eqnarray}
Note that the bulk matter contribution vanishes when $t\ne t_0$, 
because the matter would be proportional to 
$\delta(t-t_0)$, $\delta^2(t-t_0)$ and $\dot{\delta}(t-t_0)$.

If the following relation is satisfied, 
\begin{equation} \label{conticondi}
\left(\frac{\beta '}{\beta}\right)^{\dot{}}=
\frac{\dot{b}}{b}\frac{\beta'}{\beta}, 
\end{equation} 
which leads to $G_{05}=0$, then the 5D continuity equations 
are {\it automatically} satisfied \cite{kyae},  
\begin{eqnarray}
\dot{\rho}+3\frac{\dot{\beta}}{\beta}(\rho+p)
+\frac{\dot{b}}{b}(\rho+p_5)=0, \\
p_5'+3\frac{\beta '}{\beta}(p_5-p)
+\frac{\beta'}{\beta}(\rho+p_5)=0,  
\end{eqnarray} 
We will see that Eq.~(\ref{conticondi}) is satisfied for our solution.

The `$t$' dependence of $b$ in Eq.~(\ref{b(t)}) gives also 
the `$t$' dependence of the four form field, 
$H^{MNPQ}\equiv \epsilon^{MNPQR}\partial_{R}\sigma/\sqrt{-g}$, where
$\sigma$ is  
\begin{equation}
\sigma(|y|,t)=\frac{a}{4k^2}\sqrt{2A}~b(t)~ \sinh
\left(4kb(t)(|y|+y_1)\right),    
\end{equation}
where $A$ is the constant.
The `$t$' dependence of $\sigma$ generates also 
non-zero $H^{ijk5}$ component of the four form field in Eq.~(\ref{hmnpq}), 
that would be proportional also to the delta function, 
$\dot{\sigma}\propto h\propto \delta(t-t_0)$. Since the $h$ term on the 
RHS of the Einstein equations appear as $h^m/(f^2-h^2)^{n/2}$ with $0\le
m < n$, the $\delta(t)$-function of $h$ gives vanishing contribution
at $t=0$. Thus, it turns out that $T_{05}=0$ for any $t$.
For $t\ne t_0$, Eq.~(\ref{beta(t)}) satisfies the Einstein equations
with the $f^2$ dependence on the RHS~\cite{kklcc}.
Near $t=t_0$, $T_{MN}$ are reduced to those of the RS~\cite{rs2} due 
to the divergence of $\dot{\sigma}$ or $h$ (which kills the $f^2$
dependence), and so the solution $\beta$ becomes the RS 
solution~\cite{rs2}. 

We have shown above three solutions: the flat space solution
with $c_1=\tanh^{-1} (\Lambda_{old}/\sqrt{-6\Lambda_b})$ for
$t<t_0$, the RS solution~\cite{rs2} with the fine-tuning
$k=k_1$, and the flat space solution with 
$c_2=\tanh^{-1} (\Lambda_{new}/\sqrt{-6\Lambda_b})$ for $t>t_0$.
The fine-tuning solution at $t=0$ means simply that our solution
goes through such an intermediate stage. Namely, to satisfy the
field equation and the Einstein equations, $\Lambda_{old}$ suddenly 
jumps to $\sqrt{-6\Lambda_b}$ and again suddenly jumps to $\Lambda_{new}$.
If the transition from $\Lambda_{old}$ to $\Lambda_{new}$ is not
abrupt, there would exist a solution smoothly connecting the
two flat solutions with $c_1$ and $c_2$. But the above example
shows that at the intermediate stage it would go through the RS 
solution~\cite{rs2}.

The consistency of the solution is achieved by showing
Eq.~(\ref{conticondi}) using the divergence of $h$.    
As the time derivatives of the metric in $G_{55}=T_{55}+T^{(m)}_{55}$ have been 
identified already with $p_5$ of $T^{(m)}_{55}$, the remnant reads as  
\begin{equation}
\frac{\beta '}{\beta}=\pm b(t)\sqrt{-\frac{\Lambda_b}{6}
-\frac{1}{3}\frac{1}{f^2-h^2}+\frac{2}{3}\frac{h^2}{(f^2-h^2)^2}},
\end{equation}
from which we can see immediately 
that Eq.~(\ref{conticondi}) is satisfied,  
\begin{eqnarray}
\bigg(\frac{\beta '}{\beta}\bigg)^{\dot{}}
&=&\pm\frac{\dot{b}}{b}b\sqrt{-\frac{\Lambda_b}{6}
-\frac{1}{3}\frac{1}{f^2-h^2}+\frac{2}{3}\frac{h^2}{(f^2-h^2)^2}} \nonumber \\
&&\pm b\frac{\partial}{\partial t}\sqrt{-\frac{\Lambda_b}{6}
-\frac{1}{3}\frac{1}{f^2-h^2}+\frac{2}{3}\frac{h^2}{(f^2-h^2)^2}} \nonumber \\
&=&\frac{\dot{b}}{b}\left(\frac{\beta '}{\beta}\right), 
\end{eqnarray}
where we used 
\begin{equation} \label{deltadot}
\frac{\partial}{\partial t}\left[\frac{1}{\delta^2(t-t_0)}\right]=0.
\end{equation}
Note that Eq.~(\ref{deltadot}) can be shown for a specific representation
of the delta function.

We should check whether the equation of motion for the four form field 
Eq.~(\ref{fieldeq}) remains satisfied even when 
the constant $b$ is changed to $b(t)$ like Eq.~(\ref{b(t)}).  
But we can also see easily that even in the case, 
it is satisfied always regardless of $t=t_0$ or not.  
When $t\ne t_0$, it is the static case, and so the equation of motion is 
trivially satisfied. On the other hand, for $t\approx t_0$ 
the equation of motion is also satisfied due to the 
$\delta(t-t_0)$ divergence of $h$, which is checked easily 
if we use Eq.~(\ref{deltadot}).    

We have seen that there exists a time-dependent solution when
the brane tension changes from $\Lambda_{old}$ to $\Lambda_{new}$
instantaneously. In this case, the universe starting from a flat 
universe can go to a flat universe instantaneously.  
Instead of the sudden shift of the brane tension, the transition from
$\Lambda_{old}$ to $\Lambda_{new}$ can be smooth. We argue that in 
this case also, there exists a solution since we can replace the
$\delta$ function with a smooth function $D(t;\epsilon)$ of $t$ defined 
in a short interval $\epsilon$, such that 
$\lim_{\epsilon\rightarrow 0}D(t;\epsilon)=\delta(t)$. 
The error by introducing $\epsilon$ is at most $\epsilon$ and hence our 
solution is approximate, but this shows that there can exists a 
smooth solution very close to our approximate solution with an error
of order $\epsilon$.

There may exist solutions connecting de Sitter space and flat space,
de Sitter space and de Sitter space, etc.~\cite{horowitz}. 
But the final universe may be chosen probabilistically to be the flat 
universe {\it $\acute a$ la} Hawking~\cite{hawking}.

\section{Conclusion}

We considered the cosmological constant in the ($4+1$)-dimension
with a warp factor. A brane, containing the matter fields, is
located at the origin $y=0$ of the extra dimension. To separate
the 4D space we introduced a three index antisymmetric tensor
field with the field strength $H$. The simple form for the
$H$ action is considered: the Lagrangian with $1/H^2$ and $H^2$.
We found a self-tuning solution, Eq.~(\ref{solution}), with
$1/H^2$.  The $H^2$ term does not allow a self-tuning solution,
but allows nonlocalizable gravity with one brane or 
one-fine-tuning solutions with two branes.

We concentrated the discussion related to the self-tuning solution,
Eq.~(\ref{solution}). We also found that there exist de Sitter
space and anti de Sitter space solutions with $1/H^2$ term. For 
a finite range of parameters of the bulk cosmological 
constant $\Lambda_b$ and the brane tension $\Lambda_1$,
there always exists the flat space solution but it is not unique. 
Therefore, when the boundary condition is changed at the
electroweak or QCD phase transitions, the cosmological constant
after the phase transitions can be anything allowed by the
Einstein equations. However, the probability to choose the
vanishing cosmological constant is infinitely large compared
to the others'~\cite{hawking}. This argument is applicable in our
case since there exists the self-tuning solution, Eq.~(\ref{solution}).

This may not sound so attractive as any model not allowing 
de Sitter or anti de Sitter space solutions. However, if there
exists a such model, then it may not allow $\lq$inflation' which
is probably needed in cosmology.  In the self-tuning model
the transition from one integration constant to another
integration constant is possible through satisfying 
the equations of motion.

For the self-tuning solution we presented to work in cosmology,
there must be a natural explanation of the current 
acceleration of the expansion~\cite{perl}. There appeared some 
proposals~\cite{quint}, but these must be shown to work with the 
self-tuning solution.

\acknowledgments
This work is supported in part by the BK21 program of Ministry 
of Education, Korea Research Foundation Grant No. KRF-2000-015-DP0072, 
CTP Research Fund of Seoul National University,
and by the Center for High Energy Physics(CHEP) of
Kyungpook National University. 


\end{document}

%% file: jek.tex
\newcommand{\newc}{\newcommand}
\newc\eg{{\it {e.g.}}}  \newc\etal{{\it {et al.}}} \newc\ie{{\it i.e.}}
\newc\etc{{\it {etc}}} 

\newc\second{{\rm sec}} 

\newc\mone{M_1} 

\newc{\logsoms}{\log\left(s/m_{\rm eff}^2\right)}

\newc\alphas{\alpha_s}

\newc{\gstar}{g_\ast}           \newc{\gsstar}{g_{s\ast}}
\newc{\geff}{g_{\rm eff}}

\newc\meff{m_{\rm eff}} 

\newc\sigmabar{{\overline{\sigma}}}

\newc{\xf}{x_f}

\newc{\nspin}{n_{\rm spin}}
\newc{\nflavor}{n_{\rm F}}

\newc\vrel{v_{\rm rel}}

\newcommand\axino{\widetilde{a}}

\newc{\cachigamma}{C_{a\chi\gamma}}     \newc{\cachigammai}{C_{a\chi\gamma,i}}

\newc{\caww}{C_{aWW}}                   \newc{\cayy}{C_{aYY}}
\newc{\cagg}{C_{aFF}}                   

\newc{\sthw}{\sin\theta_W}              \newc{\cthw}{\cos\theta_W}
\newc{\bino}{\widetilde B}              \newc{\wino}{\widetilde W_3}
\newc{\higgsinob}{{\widetilde H}^0_b}   \newc{\higgsinot}{{\widetilde H}^0_t}

\newc{\naxino}{n_{\tilde a}}
\newc{\ngamma}{n_\gamma}

\newc{\ychi}{Y_{\chi}}                  \newc{\yeqchi}{Y^{\rm EQ}_{\chi}}

\newc{\yaxino}{Y_{\axino}}
\newc{\yeqaxino}{Y^{\rm EQ}_{\axino}}
\newc{\ythaxino}{Y^{\rm TP}_{\axino}}
\newc{\ynthaxino}{Y^{\rm NTP}_{\axino}}
\newc{\yascat}{Y^{\rm scat}_{i,j}}      \newc{\yadec}{Y^{\rm dec}_{i}}

\newc{\abund}{\Omega h^2}
\newc{\abundchi}{\Omega_\chi h^2}
\newc{\rhocrit}{\rho_{crit}}
\newc{\rhochi}{\rho_{\chi}}
\newc{\rhoaxino}{\rho_{\axino}}
\newc{\rhonu}{\rho_{\nu}}

\newc{\deltannu}{\delta N_{\nu}}

\newc{\ms}{M_{\rm S}}
\newc{\msusy}{M_{\rm SUSY}}
\newc{\mplanck}{M_{\rm P}}      \newc{\mpl}{M_{\rm Pl}}

\newc{\jxf}{J({\xf})}
\newc{\VEV}[1]{\langle #1 \rangle}

\newc{\ra}{\rightarrow}
\newc{\beq}{\begin{equation}}
\newc{\eeq}{\end{equation}}
\newc{\bea}{\begin{eqnarray}}
\newc{\eea}{\end{eqnarray}}

\newcommand\lsim{\mathrel{\rlap{\lower4pt\hbox{\hskip1pt$\sim$}}
    \raise1pt\hbox{$<$}}}
\newcommand\gsim{\mathrel{\rlap{\lower4pt\hbox{\hskip1pt$\sim$}}
    \raise1pt\hbox{$>$}}}